\documentclass[twocolumn,prl,preprintnumbers,superscriptaddress,amsmath,amssymb,english]{revtex4}
\usepackage{amsmath,amssymb,amsfonts,mathrsfs}
\usepackage{amsthm}
\usepackage{CJK}
\usepackage{bm}
\usepackage{babel}
\usepackage{graphicx}
\allowdisplaybreaks
\usepackage{braket}
\usepackage[breaklinks]{hyperref}
\usepackage{color}
\usepackage{epstopdf}
\hypersetup{colorlinks=true, linkcolor=blue, citecolor=blue, filecolor=blue, urlcolor=blue}
\UseRawInputEncoding

\begin{document}

\title{Tailoring Quadrupole Topological Insulators with Periodic Driving and Disorder}

\author{Zhen Ning}
\affiliation{Institute for Structure and Function $\&$ Department of Physics, Chongqing University, Chongqing 400044, People's Republic of China}
\affiliation{Chongqing Key Laboratory for Strongly Coupled Physics, Chongqing University, Chongqing 400044, People's Republic of China}

\author{Bo Fu}
\affiliation{Department of Physics, The University of Hong Kong, Pokfulam Road, Hong Kong, China}

\author{Dong-Hui Xu}
\email[]{donghuixu@hubu.edu.cn}
\affiliation{Department of Physics, Hubei University, Wuhan 430062, People's Republic of China}

\author{Rui Wang}
\email[]{rcwang@cqu.edu.cn}
\affiliation{Institute for Structure and Function $\&$ Department of Physics, Chongqing University, Chongqing 400044, People's Republic of China}
\affiliation{Chongqing Key Laboratory for Strongly Coupled Physics, Chongqing 400044, People's Republic of China}
\affiliation{Center of Quantum Materials and Devices, Chongqing University, Chongqing 400044, People's Republic of China}

\date{\today}

\begin{abstract}
The quadrupole topological insulator (QTI) has attracted intense studies as a prototype of symmetry-protected higher-order topological phases of matter with a quantized quadrupole moment.
The realization of QTIs has been reported in various static settings with periodic structures. Here, we theoretically investigate topological phase transitions and establish the QTI phase in a periodically driven system with disorder. In the clean limit, the Floquet QTI phase emerges from a topologically trivial band structure driven by elliptically polarized irradiation. More strikingly, starting from a pure and static system with trivial topology, we unveil an intriguing QTI phase which necessitates the simultaneous presence of disorder and periodic driving. Furthermore, we reveal that particle-hole symmetry is sufficient to protect the QTI.
Our work not only establishes a new strategy to design QTIs but also enriches the symmetry-protected mechanism of higher-order topology.
\end{abstract}
\maketitle

\emph{{\color{magenta}Introduction.}}---The Berry phase~\cite{berry1984quantal,Zak1989prl} lies at the heart of the modern theory of the macroscopic electric polarization~\cite{Vanderbilt1993prb,Reta1994rmp} in a crystalline insulator. The quantization of electric polarization based on the Berry phase and its generalization~\cite{TKNN1982,KMZ22005prl} provides a powerful language for describing and classifying the band topology of materials ~\cite{TIRMPHasan2010,TIRMP2011}. 
Recently, the extension of the quantized polarization from the dipole moment to higher multipole moments has been made~\cite{BBH,BBH2017PRB}, resulting in  the discovery of higher-order topological insulators~\cite{Langbehn2017PRL,Song2017PRL,Schindler2018SA,xie2021higher}.
Of particular importance is the two-dimensional~(2D) quadrupole topological insulator (QTI) associated with a quantized quadrupole moment, of which gapless zero-dimensional corner states feature the unconventional bulk-boundary correspondence arising from higher-order topology. Owing to intriguing topological phenomena,
higher-order topological insulators have rapidly drawn broad interest and experimentally realized in various platforms~\cite{Serra_Garcia2018Nature,Peterson2018Nature,Imhof2018NatPhys,mittal2019photonic,xue2020observation,Qi2020prl}.

Meanwhile, the exploration of topological phases has been dynamically extended to systems that are driven periodically out of equilibrium~\cite{Shirley,highf2015,Okaarcmp2019,harper2020topology,rudner2020band,bao2021light}. The periodic driving breaks the time-translation symmetry and thus leads to unique non-equilibrium Floquet engineering of topological phases~\cite{titumprx2013,maczewsky2017observation}.
Inspired by the great advancements of various Floquet topological phases in the time-dependent control, several Floquet higher-order topological phases including Floquet QTIs in periodically driven systems have been proposed~\cite{Vegaprb2019,hu2020prl,huang2020prl,c1,c2,c3,c4,c5,c6,c7,c8,c9,c10,c11,c12,c13,c14,c15,c16,c17,c18,c19,c20}.

In static topological systems, common wisdom holds that gapless boundary states are immune to weak disorder and would be destroyed by strong disorder. Surprisingly, the static disorder has been proven able to achieve nontrivial topology in some topologically trivial pure systems. A prominent example is the topological Anderson insulator~\cite{Li2009prl,groth2009prl,guo2010prl,meier2018observation,stutzer2018photonic,Liuprl2020TAI}. More recently, disorder-induced higher-order topological insulators have been found both theoretically and experimentally~\cite{li2020prl,yang2021prb,zhang2021prl}. In periodically driven systems, disorder can induce topological phases that go beyond the well-established paradigm of static disorder-induced topological phases~\cite{titum2015prl,titum2016prx,wauters2019prl}. So far, the uncovered disorder-induced Floquet topological phases possess the ordinary first-order topology. Yet, the study on the interplay of disorder and periodic driving in determining higher-order topology is still missing.


In this work, we present a systematic characterization of an emergent QTI phase in a periodically driven system subject to disorder. In the clean limit, a Floquet QTI phase with the quantized quadrupole moment, hosting four degenerate zero-energy corner modes, can be generated by elliptically polarized irradiation. More importantly, the joint effort of periodic driving and disorder can induce a QTI phase characterized by a real space quadrupole moment. This intriguing QTI phase would not exist when either periodic driving or disorder is absent, which implies a new strategy to construct quadrupole topology in two dimensions. 
Moreover, distinguished from the QTIs protected by the crystalline symmetries or chiral symmetry~\cite{BBH2017PRB,li2020prl}, the QTI phase in periodically driven systems with disorder is only protected by particle-hole symmetry, further enriching the classification of higher-order topology. 

\begin{figure}[htb]
	\centering
	\includegraphics[width=0.45\textwidth]{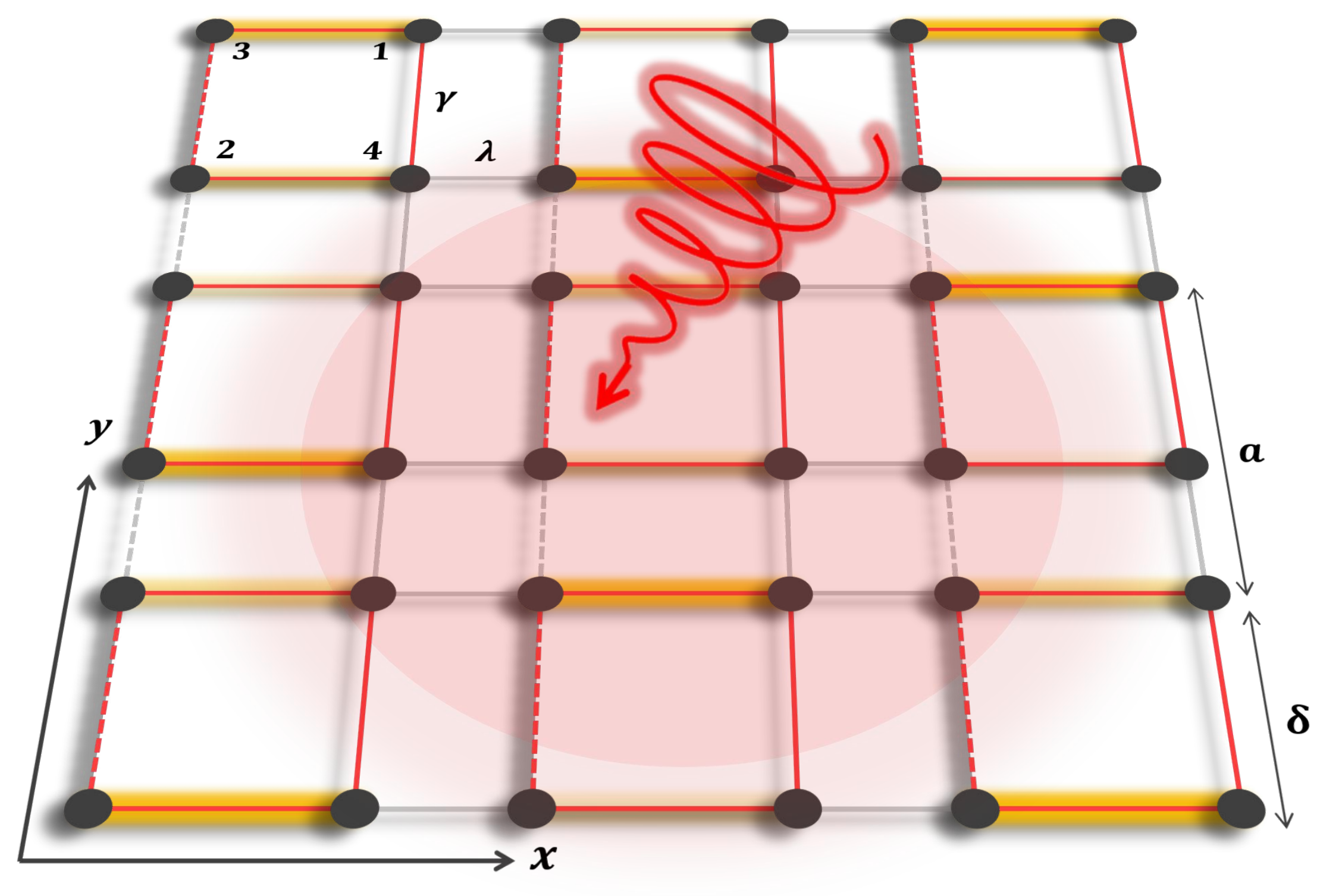}%
	\caption{ The schematic illustration of the BBH lattice model with the lattice constant $a$ under periodic driving. The distance between the nearest neighbor in a unit cell is denoted by $\delta$. The incident light is circularly polarized with $\mathbf{A}(t) = [A_x\cos(\omega t),A_y\sin(\omega t)]$. The $\gamma$ and $\lambda$ denote intra- and inter-cell hopping amplitudes, respectively. The dashed lines represent negative hopping terms.  The random halation denotes the disorder of the intra-unit cell hopping.}
	\label{Fig:BBH}
\end{figure}
\emph{{\color{magenta}Model and theory.}}---Our starting point is a 2D system described by the Benalcazar-Bernevig-Hughes~(BBH) model~\cite{BBH}, a paradigmatic model of QTIs. The tight-binding lattice Hamiltonian is given by
\begin{equation}
\begin{aligned}
H_q&=\sum_{\mathbf{r}}[\psi^{\dag}_{\mathbf{r}} T_{0}\psi_{\mathbf{r}}+
\psi^{\dag}_{\mathbf{r}+\mathbf{e}_{x}} T_{x}\psi_{\mathbf{r}}+\psi^{\dag}_{\mathbf{r}+\mathbf{e}_{y}} T_{y}\psi_{\mathbf{r}}+h.c.],\\
\end{aligned}
\end{equation}
with $T_{0} = \gamma\tau_1\sigma_0-\gamma\tau_2\sigma_2$, $T_{x} = -\frac{\lambda}{2i}\tau_2\sigma_3+\frac{\lambda}{2}\tau_1\sigma_0$ and
$T_{y} = -\frac{\lambda}{2i}\tau_2\sigma_1-\frac{\lambda}{2}\tau_2\sigma_2$,
where $\sigma_i$ and $\tau_i$ are Pauli matrices acting on different subspace within a unit cell. The
$\psi^{\dag}_{\mathbf{r}} = [c^{\dag}_{\mathbf{r}1},c^{\dag}_{\mathbf{r}2},c^{\dag}_{\mathbf{r}3},c^{\dag}_{\mathbf{r}4}]$ and $\psi_{\mathbf{r}}$
are four component spinors with creation (annihilation) operators $c^{\dag}_{\mathbf{r}\xi}$ ($c_{\mathbf{r}\xi}$) at unit cell $\mathbf{r}$ which consists of four sublattices marked by the subscript $\xi=1,2,3,4$. The $\mathbf{e}_{x,y}$ are unit vectors along the $x$, $y$ directions.
The schematic illustration of the hopping matrices $T_{0,x,y}$ is shown in Fig. \ref{Fig:BBH}. The intra- and inter-cell hopping amplitudes are $\gamma$ and $\lambda$, respectively. The topological phase transition is determined by the ratio of $\gamma / \lambda$. Without loss of generality, we will set $\lambda=a=\hbar=1$ hereafter.


We consider the system subject to a space-homogeneous periodic-driving field characterized by the vector potential $\mathbf{A}(t) = [A_x\cos(\omega t),A_y\sin(\omega t)]$ with $\omega$ being the frequency related to one period $T = 2\pi/\omega$. $\mathbf{A}(t)$ endows the hopping amplitudes of electrons an additional phase factor through the Peierls substitution $e^{i\phi_{ij}} =e^ {ie\mathbf{A} \cdot (\mathbf{x}_i-\mathbf{x}_j)}$. Now, the hopping matrices $T_{0,x,y}$ become
\begin{equation}
\begin{aligned}
T_{0} &\rightarrow
\begin{pmatrix}
0 & 0 &  \gamma e^{-i\phi^{\delta}_x(t)}& \gamma e^{-i\phi^{\delta}_y(t)}\\
0 & 0 & -\gamma e^{i\phi^{\delta}_y(t)} & \gamma e^{ i\phi^{\delta}_x(t)}\\
\gamma e^{i\phi^{\delta}_x(t)} & -\gamma e^{-i\phi^{\delta}_y(t)} & 0& 0\\
\gamma e^{i\phi^{\delta}_y(t)} &  \gamma e^{-i\phi^{\delta}_x(t)} & 0& 0\\
\end{pmatrix},
\end{aligned}
\label{eq:T0}
\end{equation}
\begin{equation}
\begin{aligned}
T_{x} &\rightarrow (-\frac{\lambda}{2i}\tau_2\sigma_3+\frac{\lambda}{2}\tau_1\sigma_0)e^{i\phi^{a}_x(t)},\\
T_{y} &\rightarrow (-\frac{\lambda}{2i}\tau_2\sigma_1-\frac{\lambda}{2}\tau_2\sigma_2)e^{i\phi^{a}_y(t)},
\end{aligned}
\label{eq:TXY}
\end{equation}
where the time-dependent phase factors are defined as $\phi^{\delta/a}_x(t) = u^{\delta/a}_x\cos(\omega t)$ and $\phi^{\delta/a}_y(t) = u^{\delta/a}_y\sin(\omega t)$. Here, $u^{\delta}_x = k_A\delta\cos\theta$, $u^{a}_x = k_A(a-\delta)\cos\theta$, $u^{\delta}_y = k_A\delta\sin\theta$, and $u^{a}_y = k_A(a-\delta)\sin\theta$ with $k_A = eA_0$. The amplitude $A_0 = \sqrt{A^2_x+A^2_y}$ and polarized angle $\theta = \arctan(A_y/A_x)$ are two important parameters to determine the topological features in the presence of periodic driving. According to the Floquet theory, we can transform the time-periodic Hamiltonian $H_q(t) = H_q(t + T)$ into frequency domain, and the time-independent Floquet Hamiltonian $H^F$ composed of Fourier components is $H^F_{nm} = n\delta_{nm}  + h^F_{n-m}$, where $h^F_{l}=\frac{1}{T}\int^T_0 dt H_q(t) e^{i l \omega t}$ with $l=n-m$. In the off-resonant case, electronic structures are effectively modified by the virtual photon absorption processes, and thus we obtain an effective Floquet-Bloch Hamiltonian by using the Magnus expansion~\cite{goldman}
\begin{equation}
\begin{aligned}
H^{\rm{eff}}({\mathbf{k}}) &= H^{\rm{eff}}_0({\mathbf{k}}) + \Delta H^{\rm{eff}}({\mathbf{k}}),
\end{aligned}
\label{eq:HFK}
\end{equation}
where $\mathbf{k}= (k_x,k_y)$ is the 2D wave vector, $H^{\rm{eff}}_0 = \frac{1}{T}\int^T_0 dt H_q(t)$ represents the lowest order term, and $\Delta H^{\rm{eff}} = \sum_{l\geq 1} [h^F_{l},h^F_{-l}]/(l\omega)$ denotes the higher-order correction. The full details of $H^{\rm{eff}}_0$ and $\Delta H^{\rm{eff}}$ are included in the Supplemental Material (SM) \cite{SM}. According to the effective Hamiltonian of Eq. (\ref{eq:HFK}), we can successfully control topological transitions of BBH model in the presence of a periodic driving field.

As seen from the detailed proof in the SM ~\cite{SM}, chiral symmetry $\mathcal{C}$ and time-reversal symmetry $\mathcal{T}$ are both broken under the elliptically polarized irradiation, nevertheless, the combination of these two symmetries (i.e., particle-hole symmetry $\mathcal{P} = \mathcal{C}\mathcal{T}$) is preserved. Particle-hole symmetry is represented by the action on the effective Hamiltonian as $\mathcal{P}H^{\rm{eff}}({\mathbf{k}})\mathcal{P}^{-1}=-H^{\rm{eff}}({-\mathbf{k}})$, where $\mathcal{P}=\tau_{3}\sigma_0 \mathcal{K}$ with the complex conjugation $\mathcal{K}$. In the SM \cite{SM}, we also prove that particle-hole symmetry is critical to the quantization of quadrupole moment in the presence of periodic-driving and even disorder. This further expands the knowledge that the quantization of quadrupole moment in static disordered systems is protected by chiral symmetry \cite{li2020prl}.





\emph{{\color{magenta}Floquet QTI.}}---We employ the winding number $n^w_{x,y}$ along the $x$ and $y$ directions~\cite{BBH2017PRB} to characterize the topology of the Floquet Hamiltonian Eq.~(\ref{eq:HFK}) as it can be mapped into two effective Su-Schrieffer-Heeger~(SSH) models~\cite{SM,2DSSH}.
The winding numbers $n^w_{x,y}$  are both $\mathbb{Z}_2$ invariants, which encode the topology of the corresponding edges. In particular, when $n^w_{x} = n^w_{y} = 1$ (in units of $\pi$), the Floquet Hamiltonian along the two directions are both topologically nontrivial. In this case, the system is in fact a QTI phase exhibiting four zero-energy corner states. The calculation of quantized quadrupole moment is presented in Fig. S2 of the SM~\cite{SM}.
When only one of the two winding numbers is non-zero, for instance $n^w_{y} = 1$ and $n^w_{x} = 0$, the edge states exist at the boundary of the $y$ direction, but are delocalized along the $x$ direction. The system is a trivial insulator when $n^w_{x} = n^w_{y} = 0$, and there is no boundary polarization. The related wave functions are depicted in Fig. S1 \cite{SM}, consistent with our winding number analysis. Therefore, we can use the sum of these two winding numbers $n^w = n^w_x + n^w_y$ to characterize different phases of the driven system.


\begin{figure}[htb]
\setlength{\belowcaptionskip}{-0.20cm}
\setlength{\abovecaptionskip}{-0.10cm}
\includegraphics[scale=0.6]{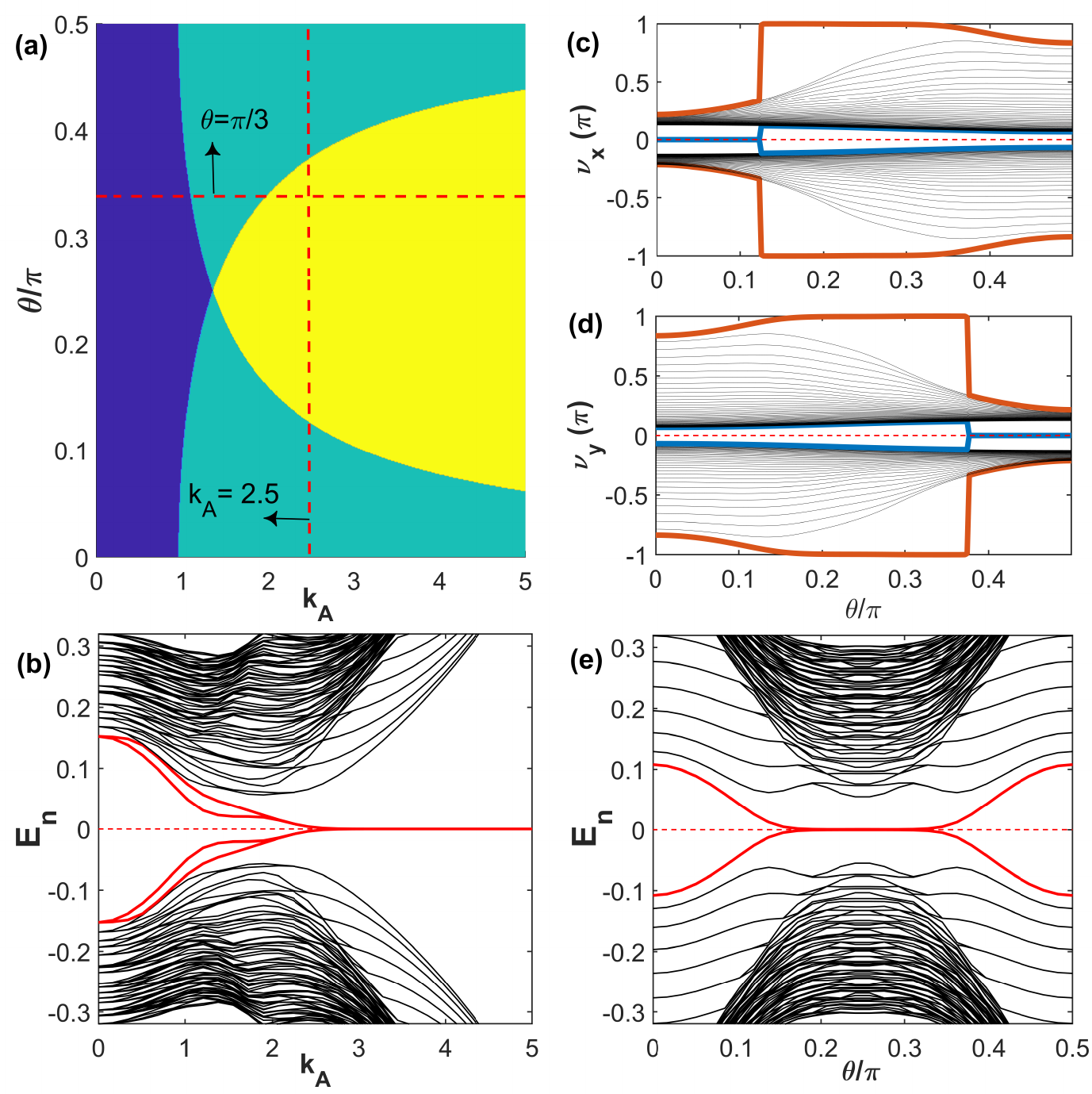}
\caption{(a) The phase diagram under periodic driving depicted by the winding number $n^w = n^w_x + n^w_y$ on the parameter plane $\theta-k_A$. Three distinct topological phases are shown, i.e., the Floquet QTI phase with $n^w=2$ (yellow), the Wannier sector topological phase with $n^w=1$ (green), and the trivial insulator phase with $n^w=0$  (blue). (b) The energy spectrum $E_n$ of Eq. (\ref{eq:HFK}) with open boundaries as a function of $k_A$ at $\theta = \pi/3$. The Wannier spectrum (c) $\nu_{x}$, (d) $\nu_{y}$ and the energy spectrum (e) $E_n$ as a function of $\theta$ at $k_A=2.5$. In panels (b)-(e), we fix the $\gamma = 1.1$.}
\label{Fig:pdiag}
\end{figure}
In the absence of periodic-driving field (i.e., $k_A = 0$), the system is in a QTI phase for $\gamma<1$ or in a trivial insulator phase for $\gamma>1$ \cite{BBH}. Next, we first reveal that the Floquet QTI phase can be created by periodic driving from a trivial insulator phase (i.e., $\gamma>1$). As shown in Fig.~\ref{Fig:pdiag}(a), the phase diagram characterized by $n^w$ in the $k_A$-$\theta$ parameter space reveals three distinct topological phases, which correspond to $n^w=0$, $n^w=1$, and $n^w=2$, respectively. The Floquet QTI phase is seen as the yellow feature with $n^w=2$. To illustrate the process of topological phase transition, we calculate the energy spectrum with open boundaries at the fixed $\theta = \pi/3$ [i.e., along the horizontal dashed line in Fig.~\ref{Fig:pdiag}(a)]. As depicted in Fig. \ref{Fig:pdiag}(b), it is found that four in-gap modes develop and merge together into zero-energy modes with the increase of strength $k_A$.

We can further calculate the Wannier spectrum to distinct the topological phases by using the Wilson loop operator~\cite{BBH2017PRB}.
For a ribbon with finite width in the $x$ direction, the corresponding Wilson loop operator is constructed as $W_{y,\mathbf{k}} = F_{y,\mathbf{k}+(N_{y}-1){\delta \mathbf{k}_{y}}}...F_{y,\mathbf{k}+{\delta \mathbf{k}_{y}}}F_{y,\mathbf{k}}$, where ${\delta \mathbf{k}_{y}} = (0,2\pi/N_{y})$ is the step length. The matrix elements of $F_{y,\mathbf{k}}$ are defined as $F^{mn}_{y,\mathbf{k}} = \langle u^m_{\mathbf{k}+{\delta \mathbf{k}_{y}}}\vert u^n_{\mathbf{k}}\rangle$, where $\vert u^n_{\mathbf{k}}\rangle$ is the $n$-th eigenvector from the occupied bands $n=1,...,N_{\mathrm{occ}}$. The Wannier spectrum $\nu_{y}$ can be obtained by diagonalizing the Wilson loop operator as
\begin{equation}
\begin{aligned}
W_{y,\mathbf{k}} = \sum_n\vert \nu^n_{y,\mathbf{k}}\rangle e^{i\nu^n_{y}}\langle \nu^n_{y,\mathbf{k}}\vert,
\end{aligned}
\label{eq:WE}
\end{equation}
where $\vert \nu_{y,\mathbf{k}}\rangle$ is the eigenvector of Wilson loop matrix. Using the similar procedure, we can obtain the Wannier spectrum $\nu_{x}$ of the ribbon with finite width in the $y$ direction.

The calculated Wannier spectra $\nu_{x,y}$ as a function of $\theta$ for a fixed strength $k_A$ [along the vertical dashed line in Fig.~\ref{Fig:pdiag}(a)] are displayed in Figs.~\ref{Fig:pdiag}(c) and \ref{Fig:pdiag}(d). We can see that the zero modes of Wannier spectrum $\nu_{x}$ (or $\nu_{y}$) are present when $n^w_y = 1$ and $n^w_x = 0$ (or $n^w_x = 1$ and $n^w_y = 0$), and thus the phase with $n^w=1$ is termed as a Wannier sector topological phase. In addition, we also find that edge modes of the Wannier spectrum are pinned to $\nu_{x} =\nu_{y}= \pi$ for the Floquet QTI phase, and its corresponding energy spectrum is plotted in Fig.~\ref{Fig:pdiag}(e). Meanwhile, it is worth noting that the Floquet QTI phase can directly be transitioned from the trivial phase when $\theta = \pi/4$ rather than undergoing the $n^w=1$ topological phase. The Floquet QTI phase is difficult to arise when $\theta$ approaches to 0 or $\pi/2$.

\emph{{\color{magenta}QTI phase driven by the interplay of periodic driving and disorder.}}---Recently, disorder in static systems has been shown to effectively change the hopping amplitude and thereby create the QTI phase from a trivial insulator phase~\cite{li2020prl,yang2021prb}. We expect intriguing higher-order topological phase transitions induced by the interplay of periodic driving and disorder in the present 2D system. To depict this process, we consider the Hamiltonian with periodic driving and disorder, which can be formally written as $\hat{H} = \hat{H}_{q}(t) + \hat{V}$. Here, $\hat{V} = \sum_{\mathbf{r}}\vert \mathbf{r}\rangle V(\mathbf{r}) \tau_1\sigma_0 \langle \mathbf{r}\vert$ describes the particle-hole symmetry preserved disorder potential [see Fig.~\ref{Fig:BBH}]. The function $V(\mathbf{r})$ takes value randomly from an interval of uniform distribution $\in[-W/2,W/2]$ with the strength of disorder $W$.
In the off-resonant regime, the Floquet bands are well separated,
and thus we can approximate the time periodic Hamiltonian $\hat{H}_{q}(t)$ as the effective Floquet Hamiltonian $H^{\mathrm{eff}}$.
For a disordered system, the approach of Wilson loop operator is no longer valid due to the translational symmetry breaking. To characterize the QTI phase in disordered systems under periodic driving, we adopt the quadrupole moment defined in real space $q_{xy} = \frac{1}{2\pi}\mathrm{Im\ \ log}Q_{xy}$ with \cite{QMR1,QMR2,QMR3}
\begin{equation}
\begin{aligned}
Q_{xy} = \mathrm{det}(\hat{U}^{\dag}\hat{Q}_{}\hat{U})\sqrt{\det(\hat{Q}_{}^{\dag})},
\end{aligned}
\label{eq:qxy}
\end{equation}
where $\hat{Q} =e^{i2\pi \hat{x}\hat{y}/(L_xL_y)}$ with the sizes of the sample $L_x$ and $L_y$, and $\hat{x}$ and $\hat{y}$ are the position operators. The unitary matrix $\hat{U}$ is constructed by the eigenvectors of the occupied energy bands $\hat{U}= [\vert u_1\rangle,\vert u_2\rangle,...,\vert u_{N_{\mathrm{occ}}}\rangle]$. When the periodic driving and disorder are simultaneously present, all crystalline symmetries and chiral symmetry are destroyed~\cite{li2020prl, SM}. However, the preserved particle-hole symmetry can protect the quantization of quadrupole moment~\cite{SM}. The established topological invariant $q_{xy}$ allow us to investigate topological phase transitions in the presence of periodic driving and disorder.

\begin{figure}[htb]
	\includegraphics[height=8.0cm,width=8.6cm]{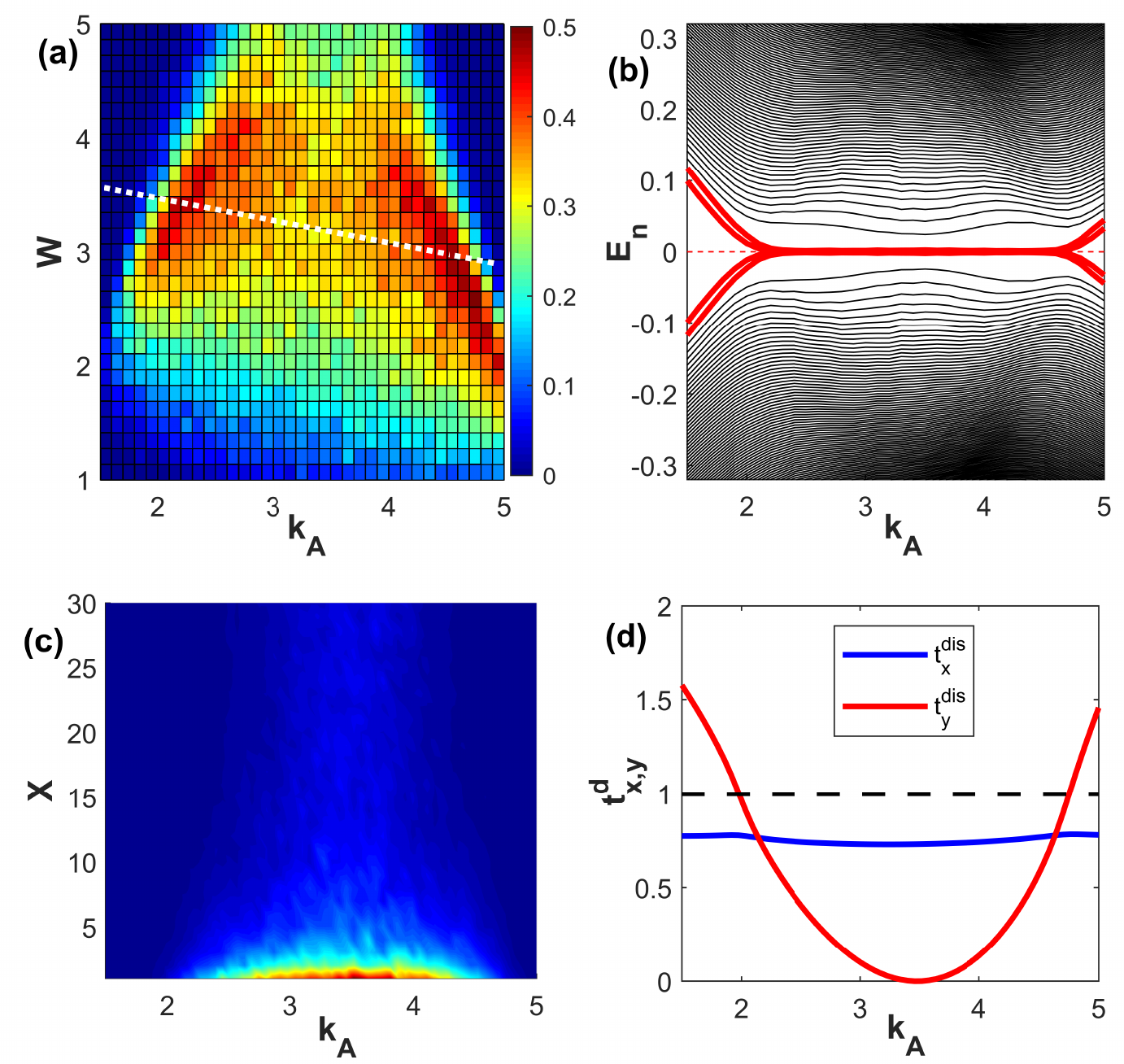}
	\caption{ (a) The phase diagram on the $W-k_A$ parameter plane depicted by the average value of quadrupole moment $q_{xy}$. (b) The energy spectrum $E_n$ with open boundaries as a function of $k_A$. (c) The local density of states $\rho(x,y=0)$ as a function of $k_A$. (d) The effective hopping amplitudes $t^{\rm{d}}_{x,y}$ as a function of $k_A$. In panels (b)-(d), we set $W = -\frac{1}{6}(k_A-5) + 3$, i.e., the dashed line in panel (a). The initial hopping parameter is $\gamma = 1.1$. }
	\label{Fig:FHOTAI}
\end{figure}

To reveal the topological phase transitions in the periodically driving system with disorder, we start from a trivial insulator phase in the clean and static limit. Based on Eq. (\ref{eq:qxy}), we obtain the phase diagram of the quadrupole moment $q_{xy}$ on the $W-k_A$ parameter plane [see Fig. \ref{Fig:FHOTAI}(a)]. Significantly, there is an explicit area of the QTI phase, which is created by the joint effort of periodic driving and disorder. This QTI phase can neither be created by the driving field without disorder, nor disorder in the absence of driving field. Notice that the averaged value of $q_{xy}$ over disorder configurations may not be exactly quantized to $\frac{1}{2}$ due to the finite size effect or limited numbers of disorder samples in numerical computations \cite{SM}. To illustrate the characteristic features of the emergent QTI phase, we calculate the energy spectrum by directly diagonalizing the tight-binding Hamiltonian with open boundaries as shown in Fig.~\ref{Fig:FHOTAI}(b). To be specific, we choose an arbitrary path in the phase diagram, for instance, the dashed white line with the disorder strength $W = -\frac{1}{6}(k_A-5) + 3$ in Fig.~\ref{Fig:FHOTAI}(a). As evident in Fig.~\ref{Fig:FHOTAI}(b), four in-gap modes at $E=0$ emerge as a function of $k_A$, implying that the presence of topologically nontrivial QTI phase. The zero-energy modes in bulk gap correspond to corner states. Therefore, we further calculate the local density of states (LDOS) $\rho(x,y)$ using the Lanczos method \cite{lanczos}. As seen in Fig.~\ref{Fig:FHOTAI}(c), the plot of LDOS $\rho(x,y=0)$ as a function of $k_A$ shows that the corner states are present in the nontrivial regime, consistent with the calculated results of $q_{xy}$.

Furthermore, we demonstrate that the topological phase transitions induced by the interplay of periodic driving and disorder can be understood by a simple picture based on the effective medium theory. To reveal this, we employ the self-consistent Born approximation (SCBA) \cite{li2020prl,yang2021prb,Li2009prl}, and the interaction between electrons and disorder is encoded by the self-energy as
\begin{equation}
\begin{aligned}
\Sigma(\epsilon) &= \frac{W^2}{12}\int\frac{d^2\mathbf{k}}{(2\pi)^2}\frac{1}{\epsilon +i0^{+} - H^{\rm{eff}}({\mathbf{k}}) - \Sigma(\epsilon)}.
\end{aligned}
\label{eq:selfE}
\end{equation}
As shown in the SM~\cite{SM}, we demonstrate the higher-order correction $\Delta H^{\rm{eff}}$ induced by the periodic driving can be considered as a perturbation in the clean limit. Hence, for simplicity, we focus on the modification of intra-cell hopping amplitudes of $H^{\rm{eff}}_0$ induced by disorder \cite{SM}.
Then, we decompose the self-energy matrix as
$\Sigma  = \Sigma_0\tau_0\sigma_0 + \Sigma_x\tau_1\sigma_0+ \Sigma_y\tau_2\sigma_2$, and the hopping amplitudes renormalized by periodic driving and disorder can be obtained from the real part of $\Sigma_{x,y}$ as
\begin{equation}
\begin{aligned}
\tilde{\gamma}_x  &=  \gamma J^{\delta}_x + \rm{Re}\Sigma_x,\\
\tilde{\gamma}_y  &=  \gamma J^{\delta}_y + \rm{Re}\Sigma_y,
\end{aligned}
\label{eq:tdis}
\end{equation}
where $J^{\delta}_{x,y}=J_0(u^{\delta}_{x,y})$ is a short-hand notation of the lowest-order Bessel function $J_0(u^{\delta}_{x,y})$, and $\gamma J^{\delta}_{x,y}$ is the photon-dressed intra-cell hopping parameters.
As displayed in Fig.~\ref{Fig:FHOTAI}(d), we plot the ratio
$t^{\rm{d}}_{x,y} = (\tilde{\gamma}_{x,y}/J^{a}_{x,y})^2$, which can characterize the QTI phase directly \cite{BBH,li2020prl}, as a function of $k_A$ along the dashed white line in Fig.~\ref{Fig:FHOTAI}(a). According to the analysis of winding numbers \cite{SM}, the topological phase transition occurs when $t^{\rm{d}}_{x} = 1$ or $t^{\rm{d}}_{y} = 1$. When $t^{\rm{d}}_{x}<1$ and $t^{\rm{d}}_{y }<1$ simultaneously, the disorder-induced Floquet QTI phase is created \cite{BBH}. This picture agrees with the numerical computations.

\emph{{\color{magenta}Summary}}--- In summary, we have theoretically presented a comprehensive study on the topological phase transitions and the emergent QTI phase in a periodically driven system with disorder. Based on the Floquet theory, we have first uncovered that a Floquet QTI phase can be generated in the clean limit by elliptically polarized irradiation. Furthermore, we have revealed that particle-hole symmetry is sufficient to protect the quantization of quadrupole moment $q_{xy}$ in the real space even though the interplay of periodic driving and disorder breaks chiral symmetry and all crystalline symmetries. The preserved particle-hole symmetry has also been confirmed by the energy spectra, as shown in Figs. \ref{Fig:pdiag}(b), \ref{Fig:pdiag}(e), and \ref{Fig:FHOTAI}(b). This well-defined higher-order topological invariant further enriches the symmetry-protected mechanism of higher-order topology. More strikingly, starting from the pure and static system with trivial topology, we have unveiled an intriguing QTI phase which necessitates the simultaneous presence of disorder and periodic driving.  Considering the recent experimental developments of topological Anderson insulators and Floquet topological insulators in photonic crystals \cite{stutzer2018photonic,zhang2021prl}, we expect the QTIs driven by periodic driving and disorder will be realized in photonic and even in condensed matter systems.

\emph{{\color{magenta}Acknowledgments.}}--- This work was supported by the National Natural Science Foundation of China (NSFC, Grants No. 11974062, No. 12074108 and No. 11704106), the Chongqing Natural Science Foundation (Grants No. cstc2019jcyj-msxmX0563), and the Beijing National Laboratory for Condensed Matter Physics.


%


\end{document}